\documentclass[aps,preprint,nofootinbib,preprintnumbers,eqsecnum,superscriptaddress]{revtex4}
\pdfoutput=1

\usepackage{wrapfig}

\usepackage{color}

\usepackage[
      colorlinks=true,
      linkcolor=blue,
      urlcolor=blue,
      filecolor=black,
      citecolor=red,
      pdfstartview=FitV,
      pdftitle={},
   pdfauthor={Stephane Detournay},
        pdfsubject={},
        pdfkeywords={},
        pdfpagemode=None,
        bookmarksopen=true
      ]{hyperref}

\usepackage[font=footnotesize,labelfont=bf,justification=centerlast,width=.94\textwidth]{caption}

\usepackage[normalem]{ulem}
\usepackage{amsmath}
\usepackage{enumerate}
\usepackage{amsfonts}
\usepackage{yfonts}

\usepackage{subfigure}
\usepackage{psfrag}

\usepackage{epsfig}
\usepackage[latin1]{inputenc}
\usepackage{float}
\usepackage{graphicx}
\usepackage{cancel}
\usepackage{mathrsfs}
\usepackage{amssymb}
\usepackage{amsfonts}
\usepackage{amsmath}
\usepackage{slashed}

\usepackage{graphicx}
\usepackage{bm}

\newcommand{\be}{\begin{equation}}
\newcommand{\ee}{\end{equation}}
\newcommand{\bea}{\begin{eqnarray}}
\newcommand{\eea}{\end{eqnarray}}

\begin{document}

\title {The rubber band and the ruler}

\author{Jean-Marie Fr\`{e}re}
\affiliation{Physique  Th\'{e}orique, CP 225, Universit\'{e} Libre de Bruxelles, B-1050 Bruxelles, Belgium}

\begin{abstract}
\vskip 0.3in
In this short presentation, we address two somewhat separate issues. The first one deals with the establishment (vs discovery)
of what we call "physical laws". The discussion runs on a "successive approximations" approach, suited to our own thinking process,
and aiming at the "simplest \emph{ possible }formulation ", as opposed to the idea of pre-existing laws to be discovered.
In the second part, we review briefly an approach to the (observed) metric originating from a short distance fractal structure,
and remind how a symmetrical situation can arise from mere re-summations.

\end{abstract}
\begin{figure}[b]
\includegraphics[width=\textwidth]{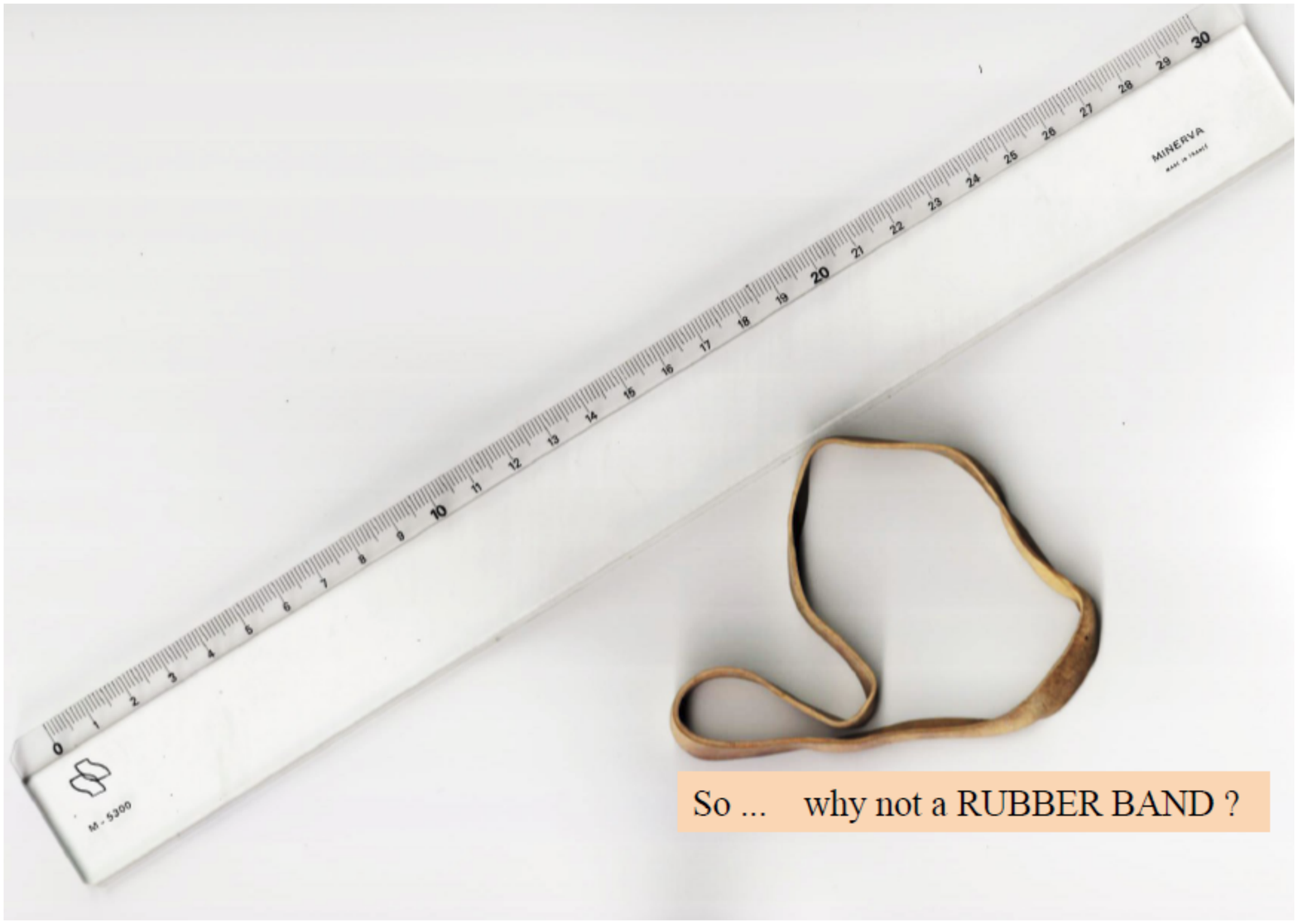}
%\caption{Length of a table in "Rubber Band" units, as a function fo time .}
\label{Rubber}
\end{figure}

\vfill

%\today

\maketitle

\tableofcontents

\section{How to choose a good ruler}
In this part, I would like to ask the reader to make abstraction of any knowledge about what might be a "good" ruler,
in particular, modern approaches like the wavelength of a particular atomic transition or more antiquated ones like some piece of metal located close to Paris. The purpose here is one of example. The obvious point to make (and this came in discussions with Fran\c{c}ois Englert when we were preparing a freshman physics course) is that it is impossible to say a priori which choice of ruler is best.
To put things in a nutshell, we can only test the "constancy" (a desirable property for a ruler) of a particular ruler by comparison to another one!
(and even in this process, assumptions are made about the fact that the length does not change during the translations and rotations needed to bring the two candidates alongside).
If more illustration is needed, we could (from everyday experience, comparing to many different objects around us) imagine that a piece of wood would be a good ruler in a desert, where the day/night temperatures differ widely, but where the humidity is constant, while a piece of metal would constitute a " better choice" in, say, Belgium. Even measuring the temperature to compensate for such effects would rely on another "ruler",
the scale of the thermometer!
This digression was probably unneeded, but it still hints at one thing: we expect a "good ruler" to provide some consistency, when applied to a large number of independent measurements of objects we expect to be reasonably stable...Thus the result of a collection of measurements, or, as we shall see, the simplicity of the laws resulting from the choice of a  particular  ruler leads to which decide on its quality.
Still,  on which basis are we to judge the "simplicity"?

To illustrate the process in a willingly caricatural (but, I think, instructive) situation, I will take an extreme case: since no ruler is "a priori" better than another, let us choose a rubber band as our standard of length.

Armed with this new tool (or experimentalist toy), we now want to perform some measurements. In this caricatural example, I suggest that we measure the length of a table every hour over a period of, say, 2 days (in fact, this regularity is not necessary, we just need a sufficient number of measurements over a relatively long time). Due to a lack of funding (hard to understand) for this project, and a similar lack of volunteers, I will just provide here some "simulated" measurements. (Your collaboration is welcome in the future!)

\begin{figure}[h]
\includegraphics[width=\textwidth]{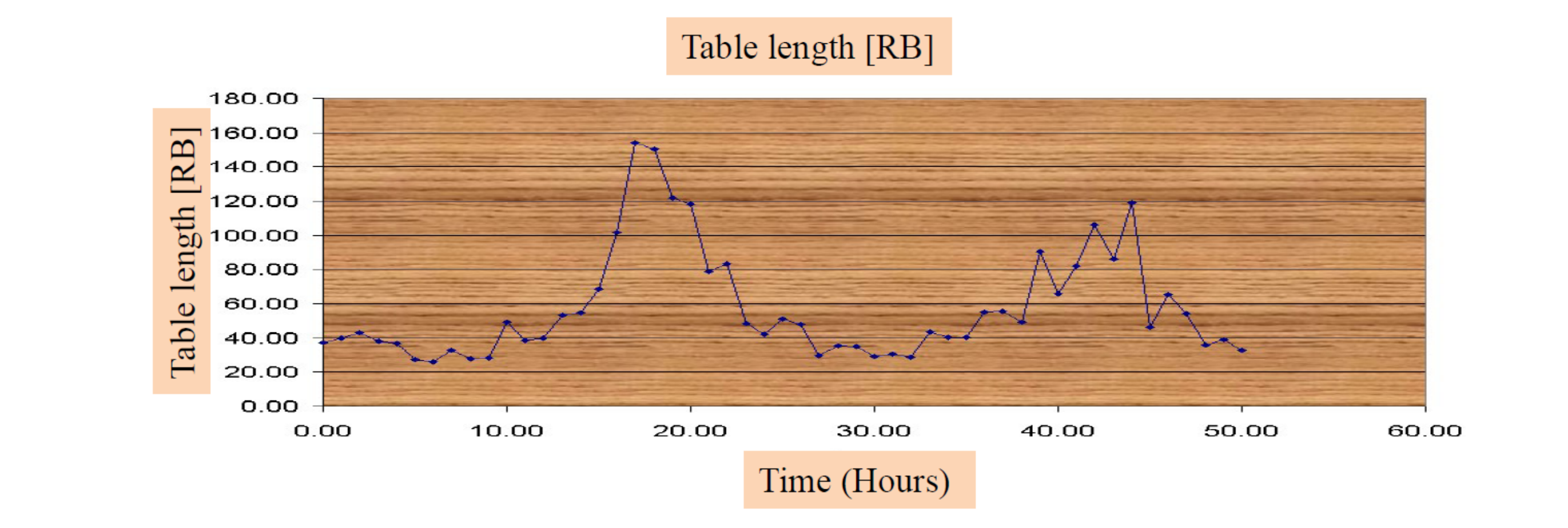}
\caption{Length of a table in "Rubber Band" units, as a function fo time .}
\label{table}
\end{figure}

We present these measurements in  Figure \ref{table}

\begin{figure}[h]
\includegraphics[width=\textwidth]{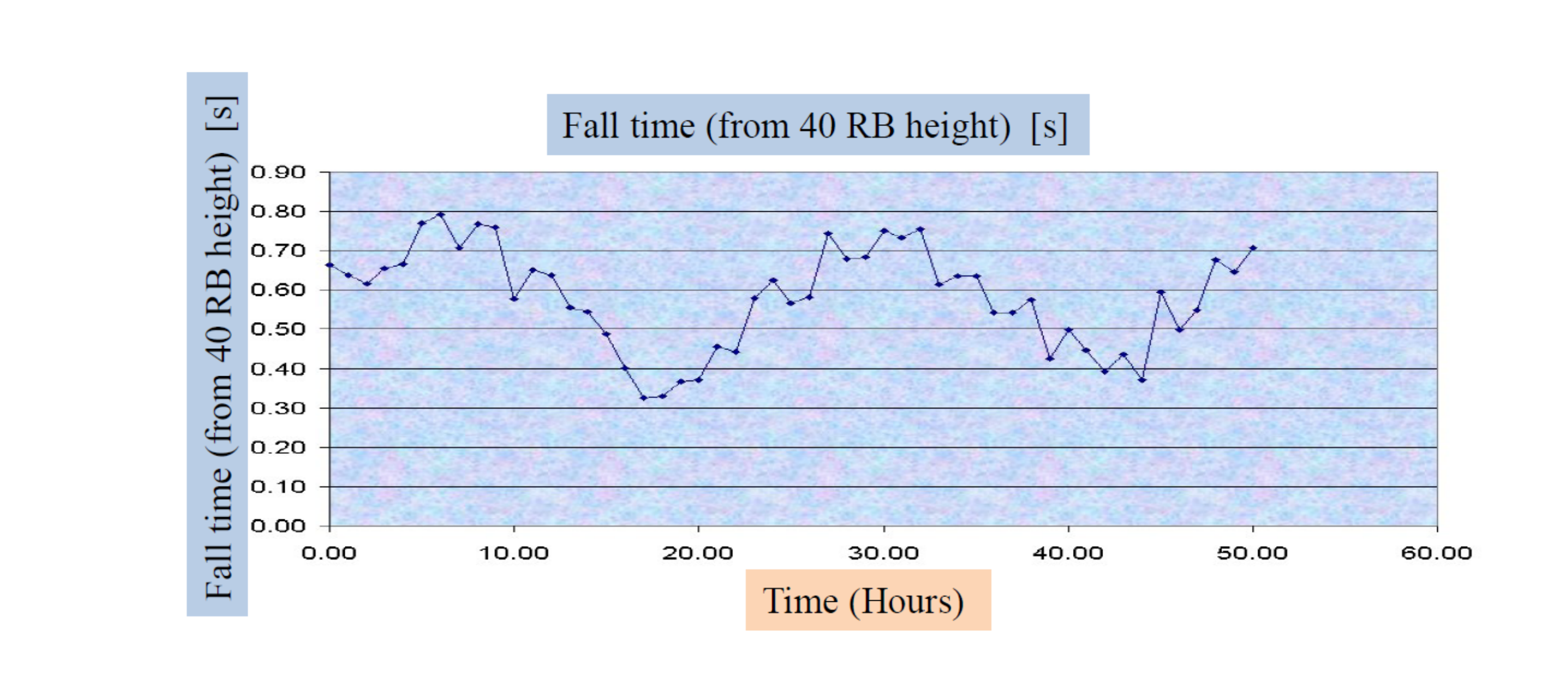}
\caption{Fall time of a marble from a height of 40 Rubber Bands, as a function of time.}
\label{marble}
\end{figure}

For a physicist, this set of data seems difficult to exploit. It may have some interest for a psychologist or zoologist, who may want to relate this to the circadian rhythms, but does not suggest the obvious predictability that we expect from a "physical law".
Fortunately, since our experimenter was rather idle between measurements we asked him/her/it (the latter applies, since we deal with virtual measurements) to perform a second experiment. Immediately after measuring the table, a marble was dropped from the exact heights of 40 Rubber Band units. We assume that no accidental errors need to be considered, and show the results in Figure  \ref{marble}; it is also assumed, for simplicity, that our experimenter has a "good" chronometer.

\begin{figure}[h]
\includegraphics[width=\textwidth]{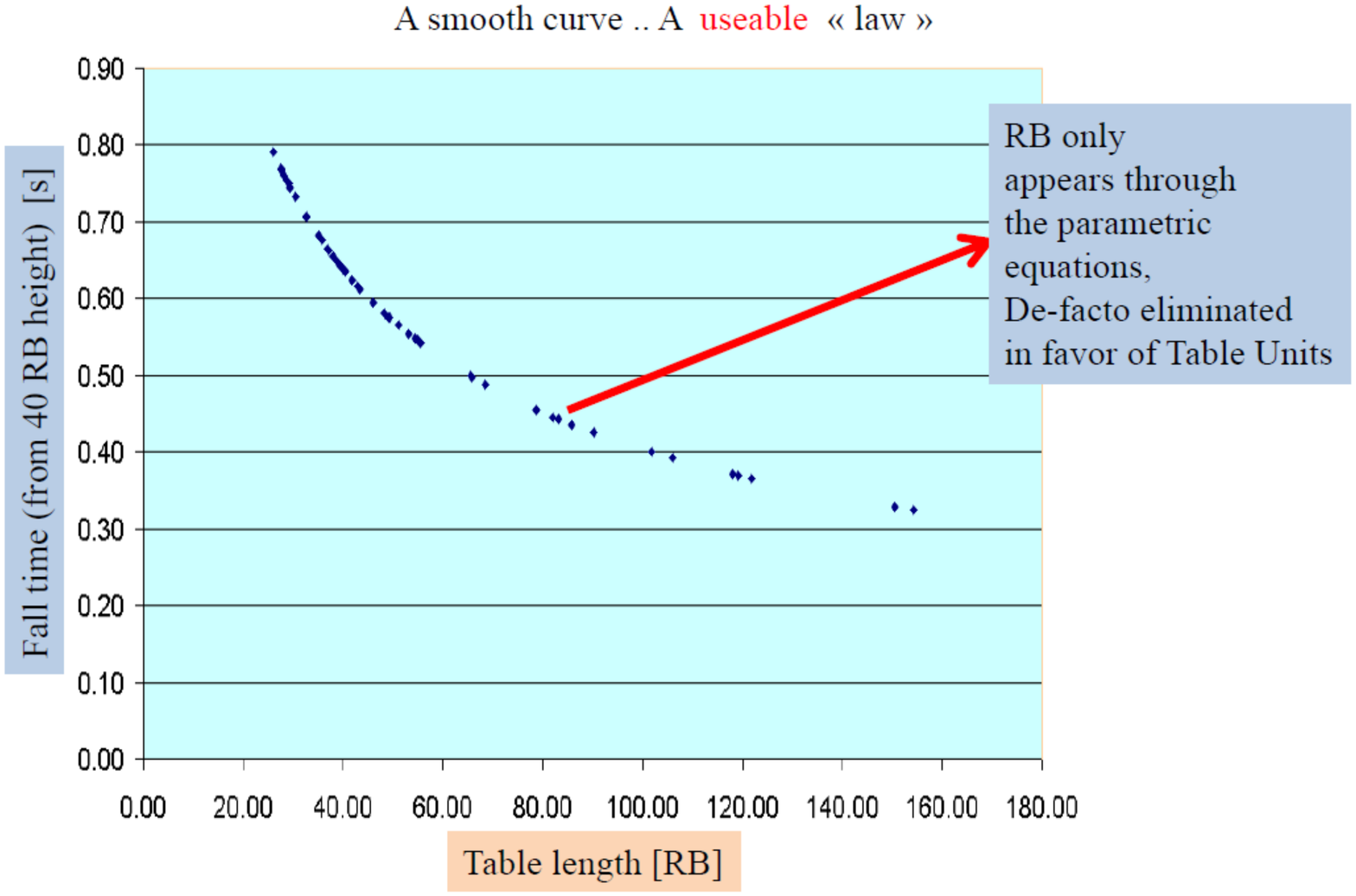}
\caption{Fall time of a marble from a height of 40 Rubber Bands, vs table length.}
\label{vs}
\end{figure}

The same comments would apply as to the usability of these data, but the trained eye of a physicist cannot fail to notice (anti-) correlations between the graphs, (the more so that we have ignored any accidental errors).
It becomes then very tempting to plot one against the other, in a graph (fall time from 40 RB vs length of table in RB), this yield figure \ref{vs}

This time, we find something close to what we might expect to call a "physical law"  (think of the many "NameOfSomeone" laws in your beginning physics course, which were just similar observations, ..). How do we recognize this? One simple answer is that it is "usable". At the difference of the previous curves, the regularity of this one suggests that interpolation would work. If given a "table length" we can predict the corresponding "fall time".
Another thing that this graph suggests is that the "rubber band" no longer appears directly, it is merely in the parametric equations of the curve, as is the time at which the measure was made, but does not figure directly in the relation. This translate the fact that it is now a simple intermediary to relate the drop height of the marble to a fraction or multiple of the table length.
It is then very tempting to just do away with the "Rubber Band units", and to use the length of the table as a primary measurement. Such a move will make full sense if a range of other measurements would indeed favour this choice.

Still it is remarkable that starting from a (rather obviously inadequate) tool, the physics itself guides us to a more suitable one.

At this point, I would like to comment on the process used. The choice of a unit of length was a pretext to discuss what we recognize as a "physical law", and in the present context, it amounts to "a law we can use to make predictions", but also a law which describes things in a "simple" way. By simple, we mean simple for us to figure out, and to use.

It would be interesting to think on how our own senses influence such a description. Certainly the importance of our vision plays a r\^{o}le (see also the "note added" below).

Another remark is that the notion of "simple" must be qualified, and trying to establish laws on the (subjective) principle of "simplicity" alone may lead to problems (think of the epicycles trap ... ). As a matter of fact, we have learnt (and the students still suffer a lot to learn) to find "simple" the simplest formulation which allows for a correct description.
As an example, consider this equation, which we all find one of the simplest possible:
$$ \Box \Psi= - m^2 \Psi$$

and think of the amount of explanation needed to explain this "simple" relation to an outsider (who might read is as "square fork = - m two fork") ...(I chose the "simplest possible" expression as a reference to the "best of all possible worlds" as criticized  by Voltaire in a satire of Leibnitz)

\subsubsection*{Note added}
At the time of writing up these proceedings, I think of a possible additional remark. Imagine that, instead of plotting 2-dimensional graphs, we had used the full machinery and made 3-dimensional representations (time of data taking, length of table, fall time). With sufficient statistics, we would have noticed, after looking at a number of 3D plots, that all points fell on a single smooth surface... would it have led to the same conclusions? Is a limited visual representation forcing us/ helping us through the search for correlations into the abstraction of laws?
Remember how the "perfect gases" law $PV=nRT$ was first taught as three separate 2-variables laws...

\subsubsection*{Inventing or discovering laws? }
The previous considerations were aimed at showing, on the basis of a caricatural example, how "physical laws" could be induced from
observation with "efficient prediction" as a guiding principle (others would call this "elegance").  Meanwhile we mentioned that this "efficiency" was probably in some way subjective, depending on our senses or thought processes (or mathematical / numerical tools available).
An opposite view would consider pre-existing "laws of Nature", which are for us to \textbf{discover}; pure mathematics being sometimes taken as such an example.

Whatever the approach (and the choice will not be clear until we have a "full description"), empirical verification (or inspiration) remains the fundamental tenet of physics. (Think of the observation of parity violation in particle physics, which was difficult to expect on the basis of "elegance", although it can be justified after the fact by the use of the smallest representations of $SL(2,C)$.

\section{The origin of a "smooth", isotropic metric}
This section is completely different from the previous one, although it shares a common point: our perception of the metric of the space surrounding us. While I may differ with Philippe Spindel's approach in the previous version, the present one evokes some work performed in common with him, together with Fran\c{c}ois Englert and Marianne Rooman.
It starts from the question: what if the microscopic space structure is discontinuous? How would we perceive it at long distances? Is the isotropy/symmetry of space a microscopic notion, or simply the result of re-summing a large number of steps to reach our observation scale?

This work was based on the idea of fractals, the reason being their self-similar structure at all scales.
For technical details, we send back to the original reference   \cite{Englert:1986tq}  and for a more sketchy presentation to   \cite{Frere:1985nm} .\begin{figure}[h]
\includegraphics[width=\textwidth]{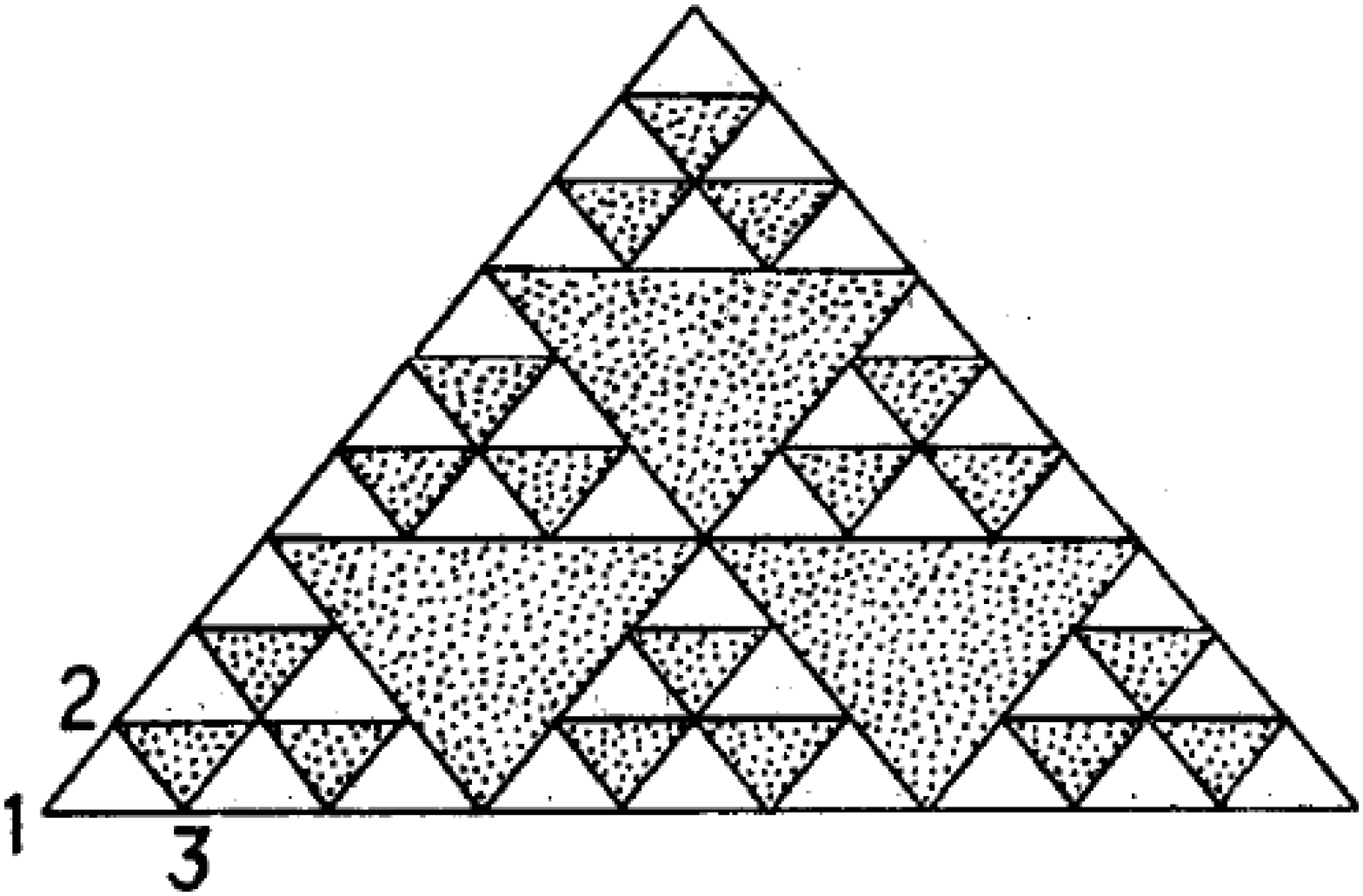}
\caption{3 successive levels of farctalisation for a triangular structure.}
\label{FractalOveral}
\end{figure}          .

Assuming some microscopic structure, the fractal is an elegant solution in that it does not really need an intrinsic scale, but rather relies on a self-similar structure at every scale.
A typical example involving 2 dimensions is the subdivision of a triangle, as shown in Figure \ref{FractalOveral}. At each level, the sides of a triangle are cut in a number of pieces (here, 3), and only the "upward-pointing" surfaces are kept. The number of subdivisions is kept the same at each level, providing thus a self-simiilar structure (after "trivial" rescaling, it is impossible to determine at which fractalisation level one is.).

On such a structure, it is possible to define a geometry, by fixing the "length" of the links. One can then, at a given fractalisation level write the equivalent of the  (discretized) Laplace equation,  (second derivatives being replaced by finite differnces).

Assuming we have placed a source charge $e$ at node 0, this leads to::
\begin{eqnarray}
  g^{ij} (\phi_i-\phi_i) &=& e \delta_{i0} \\
   g^{ij} &=& l^2_{ik} + l^2_{jk} - l^2_{ij}
\end{eqnarray}

where $l_{ik}$ is the length of link (ik).

From this equation, we now want to see whether we can easily spot the discontinuous or fractal structure of the space, through measurements.
First of all, we must realize that the structure we are dealing with, even at a single fractalisation level, differs fundamentally from a lattice.

In a lattice indeed, if we move sufficiently far from a charge, a growing number of nodes participate in the equations, 
in other tems, the "flux" emitted by the charge is spread more or less evenly on a large perimeter, 
which grows with the distance, precisely which leads in 3 dimensions to the $1/R^2$ behaviour, 
and to the vanishing of the field at infinite distance. Nothing like this here: at each step of the fractalisation, 
the entire flux is concentrated at the summits of the "large triange". In fact, it becomes impossible to define the problem 
by an isolated charge inside the triangle alone, without specifying how the flux is shared among the summits.

The first question thus is whether, at least far from the edges, the field $\phi$ behaves in a manner sufficiently close to the "continuous" one.

\begin{figure}[h]
\includegraphics[width=\textwidth]{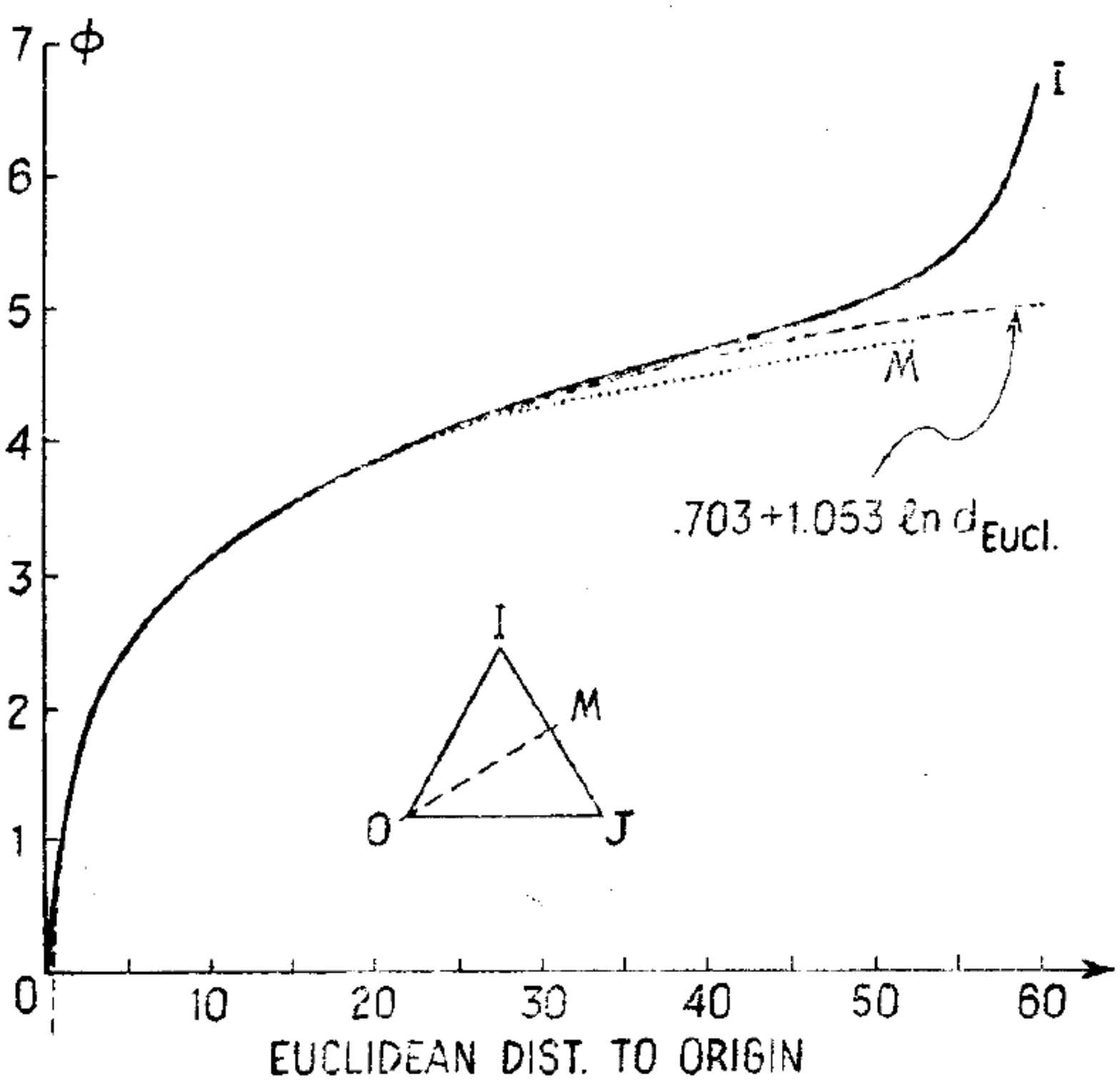}
\caption{Field inside a fractal  (in symmetrical configuration, see text).}
\label{F}
\end{figure}

We have performed this calculation for various types of subdivision, (n being the number of subdivision of a side, we had to resort to numerical calculations for larger n), and checked that indeed 
(at least for symmetrical outgoing flux -- what we could call "mirror charges"), the solution 
was close to the continuum one, as shown in Fig. \ref{F}.

          .

\subsubsection{Rescaling and Fixed points}

The link to the first part of this contribution occurs through rescaling. Assume that the metric is known at some fractalisation stage $m$, and
that we want to move one step up (to larger structure, we will call this level $m-1$). We want now to ignore the finer subdivision, and see which new metric we need to use.
How will we define the new metric? In a close analogy with the lines of the first section, we will rely on a physical phenomenon, namely the field corresponding to a local charge (think of the electric field for an electrostatic charge, or, in Minkowski space, an electromagnetic wave amplitude). By the way, this is close to the current definition of the meter, based on the length travelled by light in a unit of time!
We thus study how the coefficients $g_{ij}$ evolve from one level to the other.

For simplicity, we assume a symmetry between two of the coefficients, namely
\begin{equation}\label{scale}
    g^{1,3}=g^{2,3}=1 \neq g^{1,2}\equiv a
\end{equation}
and, at the larger stage we get  (we explicitly rescale one of the coefficients, as this measuring scale is arbitrary, and we are mainly interested in the evolution of the ratio of the lenghts):
\begin{equation}\label{scaleA}
    G^{1,3}=G^{2,3}=1 \neq G^{1,2}\equiv A
\end{equation}

For low subdivisions at least, the calculation can be easily worked out, and we obtain, for instance if the sides of the triangle are divided by 3, 
\begin{equation}\label{rescaling}
    A = \frac{11 a^4 +80 a^3 +172 a^2 + 132 a + 25}{2*(9 a^3+ 54 a^2+ 95 a+52)}
\end{equation}

The interest of this equation appears when we apply the rescaling in a recursive way, namely moving to larger and larger distances, and forgetting about the microscopic structure. In this process, we look for a fixed point, which would describe a large-scale tendency.
This is shown in figure \ref{Fixed}
\begin{figure}[h]
\includegraphics[width=\textwidth]{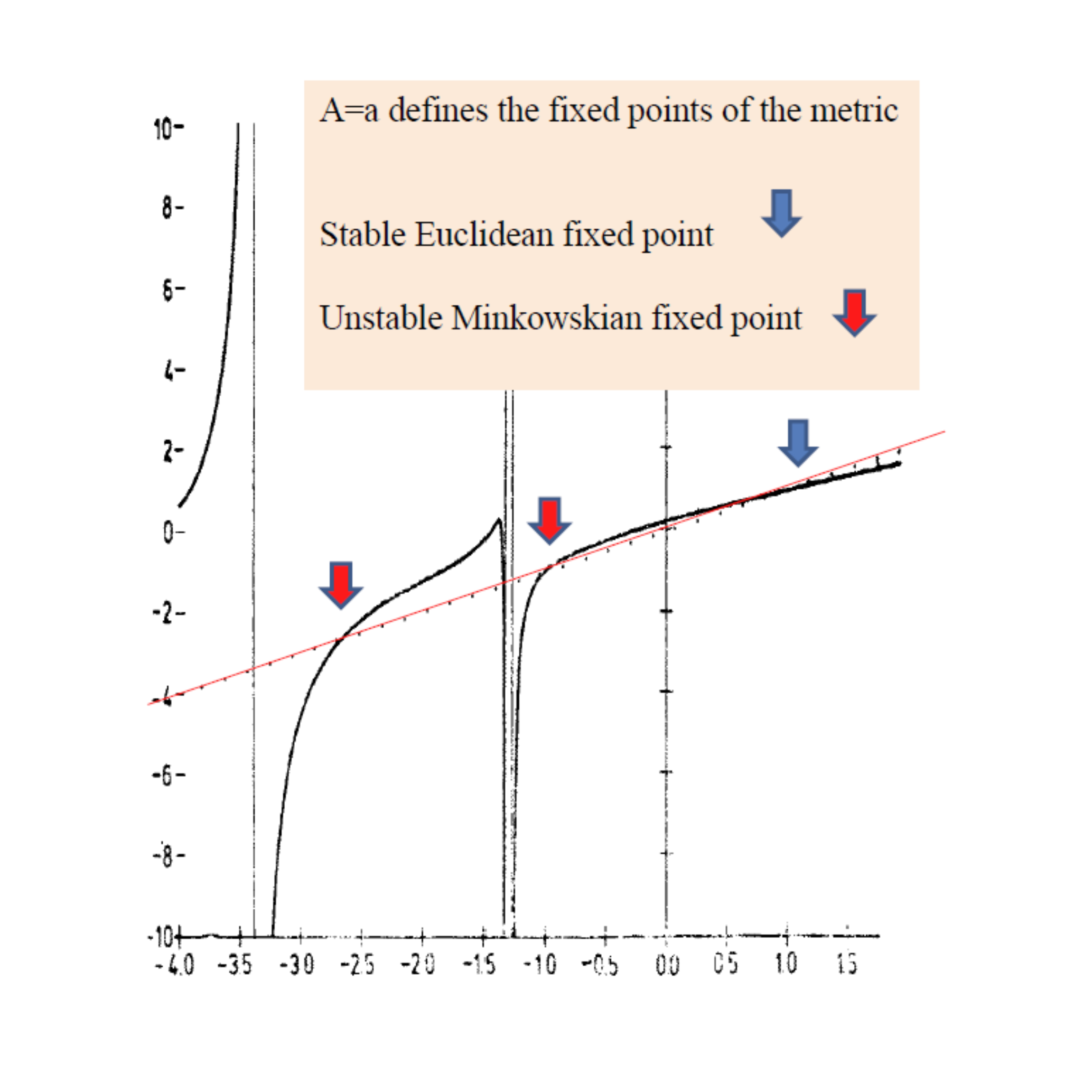}
\caption{Stable Euclidean symmetrical fixed point and unstable Minkowskian ones for n=3 subdivision and recursive rescaling.}
\label{Fixedl}
\end{figure}

From figure \cite{Fixed}, it is clear that, from an initially asymmetrical network we evolve towards a symmetrical euclidean fixed point (the minkowskian fixed points exist, but are unstable).

I cannot resist the temptation to provide an analogy (which actually can be seen as an analogical calculation) to visualize this evolution towards the larger symmetry.

Consider indeed a resistors network. We can make a complete analogy between Laplace equation (in its discrete form) and the propagation of a current, namely:
\begin{eqnarray}
  \phi_i &\rightarrow& V_i \\
  G_{ij}&=& \frac{1}{R_{ij}}\\
  I_{ij} &=&\frac{V_{ij}}{R{ij}} \\
  \sum (\phi^i - \phi^j) =0 & \rightarrow & \sum I_j = 0
\end{eqnarray}

\begin{figure}[h]
\includegraphics[width=\textwidth]{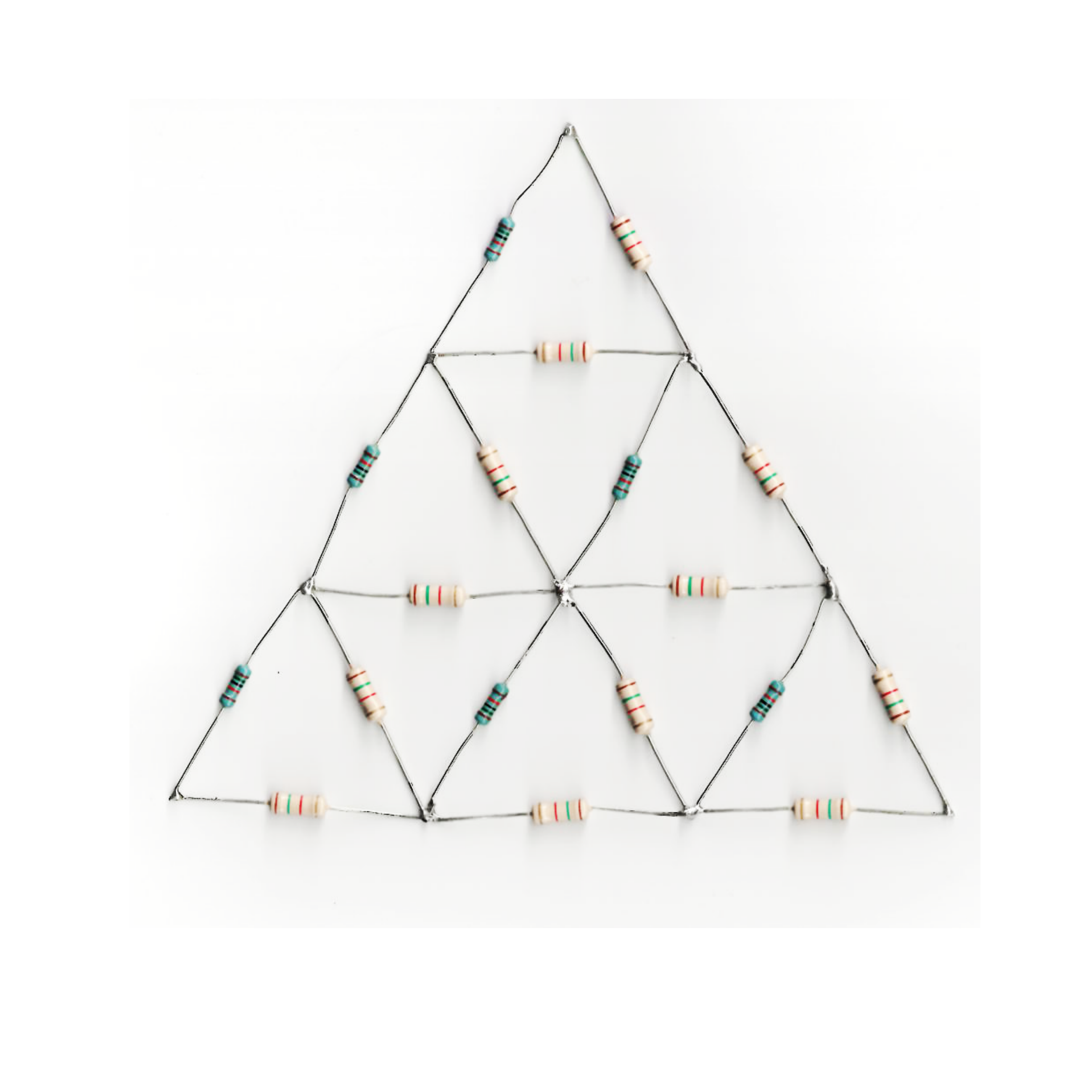}
\caption{An equivalent resistors network makes the fixed point evident.}
\label{ResistorNetworkl}
\end{figure}

The picture then becomes obvious by contemplating (or measuring) the following resistor lattice (composed of $15 k\Omega $ and $ 1.5  k\Omega$
resistors). While the horizontal path offers little resistance, and is in parallel with much larger resistance shunts, the opposite is true
for one of the other sides: an obvious large  resistance of $45  k\Omega$ is shunted by numerous lower-resistance paths.
As a result, when we look on the large scale, we tend to equalize the behaviour in the 2 directions. This argument also explains the attractive nature of the fixed point, as the process continues as long as the "direct" path offers more resistance than its alternatives.

I don't want to make conclusions here (they are quite obvious from the text), except that there is room for thought as to how much of the physical laws are defined by our measuring choices, and, ultimately, our thought process, which they in turn condition.

\providecommand{\href}[2]{#2}\begingroup\raggedright\endgroup

%\bibliography{all}

\end{document}